\documentclass[12pt]{smdqart}
\usepackage{amsfonts}
\usepackage{amssymb}
\usepackage{graphicx}
\usepackage{epsfig}
\usepackage{color}

\begin{document}

\resrep{Fat-tailed and compact random-field Ising models on cubic lattices}

\author{Nuno Crokidakis$^{1}$ and S\'{\i}lvio M. Duarte Queir\'{o}s$^{2}$}

\address{$^{1}$ Instituto de F\'{\i}sica - Universidade Federal Fluminense \\
Av. Litor\^anea s/n, 24210-340 Niter\'oi - RJ, Brazil}
\address{$^2$ Centro de F\'{\i}sica do Porto \\ Rua do Campo Alegre 687,
4169-007 Porto, Portugal}

\ead{\mailto{nuno@if.uff.br},\mailto{sdqueiro@gmail.com}}

\begin{abstract}
Using a single functional form which is able to represent several different classes of statistical distributions, we introduce a preliminary study of the ferromagnetic Ising model on the cubic lattices under the influence of non-Gaussian local external magnetic field. Specifically, depending on the value of the tail parameter, $\tau $ ($\tau < 3$), we assign a quenched random field that can be platykurtic (sub-Gaussian) or leptokurtic (fat-tailed) form. For $\tau< 5/3$, such distributions have finite standard deviation and they are either the Student-$t$ ($1< \tau< 5/3$) or the $r$-distribution ($\tau< 1$) extended to all plausible real degrees of freedom with the Gaussian being retrieved in the limit $\tau \rightarrow 1$. Otherwise, the distribution has got the same asymptotic power-law behaviour as the $\alpha$-stable L\'{e}vy distribution with $\alpha = (3 - \tau )/(\tau - 1)$. The uniform distribution is achieved in the limit $\tau \rightarrow \infty$. Our results purport the existence of ferromagnetic order at finite temperatures for all the studied values of $\tau$ with some mean-field predictions surviving in the three-dimensional case.
\end{abstract}

\keyw{Random-Field Ising Model, Fat tails, Compact distributions}


\data{20th Obtober 2010}



\maketitle

\section{Introduction and motivation}

Disordered magnetic systems have long ceased being a ``mere'' theoretical exercise to find a limelight place in the class of problems which enables the reasoning about the equilibrium and out-of-equilibrium behaviour of several compounds that can be experimentally studied. Accordingly, the Sherrington-Kirkpatrick
spin-glass model~\cite{SKmodel} and the random field Ising model (RFIM)~\cite{ma} are quintessential systems that have soon found a correspondence with condensed matter and materials science case-studies~\cite{applications}. For uncertain reasons, perhaps related to the nature of the replica method and the took for granted universality of the Gaussian, most of the studies have been devoted to the analysis of systems with $n$-modal or $n$-Gaussian disorder in either the coupling constant or the local external magnetic
field~\cite{aharony,schneiderpytte,kaufman,nuno_jpcm,octavio,berker1,hernandez,machta06}. However, the last decade has been particularly prolific in examples of the ubiquity of non-Gaussian distributions in natural systems and man-made phenomena~\cite{photons,ctbook}. Therefore, it was completely upheld the introduction of generalized disordered magnetic systems in which fat tailed and compact distributions in the local magnetic field~\cite{part1} and fat tailed ($\alpha $-stable L\'{e}vy) distributions in the coupling constant were considered~\cite{cizeaubouchaud,levyglass,mezardepl,mezard}. As a matter of fact, earlier experimental studies on organic charge-transfer compounds like N-methyl-phenazium tetra-cyanoquinodimethanide (NMP-TCNQ), quinolinium-(TCNQ)$_{2}$, acridinium-(TCNQ)$_{2}$ and phenazine-TCNQ suggested the experimental validity of non-Gaussian (fat-tailed and uniform) distributions.~\cite{fat-materials} Furthermore, in alternative quantities and systems the importance of non-Gaussian distributions has been also demonstrated. For instance, in Refs.~\cite{levyglass,mezardepl}, it is shown that for the $\alpha $-stable L\'{e}vy spin-glass the local fields are not Gaussian distributed.

Concerning the non-Gaussian RFIM, a recent work~\cite{part1} analyzed the case of distributions with power-law asymptotic behaviour, namely the generalized $r$-distribution and the Student-$t$ \cite{andrect} in the limit of infinite-range interactions. In this case, interesting results were found, namely the appearance of a change of the concavity of the critical transition line for $\tau>2$ and a corresponding divergence of the zero-temperature free energy per spin. Despite the fact that mean-field surveys are unquestionably important for a sketchy description, most of the systems are not experimentally studied nor naturally exist at a dimension equal or higher than the upper critical dimension. Consequently, you should turn our attention to systems with achievable dimensionality and worth of study, \emph{i.e.}, exhibiting critical behaviour. In this particular case, it is known that the RFIM cannot undergo a discontinuous phase transition for $d \leq 2$ because of the existence of a single minimum energy state~\cite{aizenman}, whereas Landau-like energetic arguments purport the impossibility of a ordered phase at 2D~\cite{ma}, \emph{i.e.}, no phase transition at all. Alternatively, the field-theory approach based on the existence of the dimensional reduction does not hold for the RFIM~\cite{giardina}, thus leaving the existence of critical behaviour for 3D systems under dispute.

In order to better understand this problem, we herein report preliminary results on the study of the critical frontier of the RFIM on cubic lattices with short-range (nearest-neighbour) interactions within the same quenched disorder conditions of Ref.~\cite{part1}. To that, we performed Monte Carlo (MC) simulations on lattices with sizes in the standard range. The figures obtained point at the existence of ferromagnetic order at finite temperatures for a wide range of the parameters of the random-field distribution and that in some cases continuous phase transitions exist with the critical temperature being defined by both the tail and the width of the distribution.

\section{Model and Monte Carlo Simulation}

We have considered the following Hamiltonian on a cubic lattice of linear size $L$:
\begin{equation} \label{1}
\mathcal{H}=- J\sum_{<i,j>}S_{i}S_{j} - \sum_{i}H_{i}S_{i}~,
\end{equation}
where $S_{i}=\pm 1$ and the sum $\sum_{<i,j>}$ being performed over the nearest-neighbour (NN) pairs of spins.
$J$ is the ferromagnetic interaction constant between NN and the random fields $\{H_{i}\}$ are quenched variables
ruled by a PDF that is defined by a parameter $\tau $ (generic degree of freedom). For $\tau <1$,
\begin{equation}
P_{i}(H_{i})=\sqrt{\frac{1-\tau }{\pi }\mathcal{B}_{\tau }}\frac{\;\Gamma %
\left[ \frac{5-3\tau }{2\left( 1-\tau \right) }\right] }{\;\Gamma \left[
\frac{2-\tau }{1-\tau }\right] }[1-\mathcal{B}_{\tau }(1-\tau )H_{i}^{2}]^{%
\frac{1}{1-\tau }},  \label{2}
\end{equation}%
(\noindent with $\left\vert H\right\vert \leq \left[ \mathcal{B}_{\tau
}(1-\tau )\right] ^{-1/2}$) which is the generalized $r$-distribution, and
for $\tau >1$, we have
\begin{equation}
P_{s}(H_{i})=\sqrt{\frac{\tau -1}{\pi }\mathcal{B}_{\tau }}\frac{\;\Gamma %
\left[ \frac{1}{\tau -1}\right] }{\;\Gamma \left[ \frac{3-\tau }{2\left(
\tau -1\right) }\right] }[1-\mathcal{B}_{\tau }(1-\tau )H_{i}^{2}]^{\frac{1}{%
1-\tau }},  \label{3}
\end{equation}%
\noindent which is the generalized Student-$t$ distribution also widely named $q$-Gaussian within statistical mechanics~\cite{ctbook}.
By \textit{generalized} we mean that the degrees of freedom, $m$ and $n$, of $t$- and $%
r $-distributions are extended to the entire domain of feasible real values
according to the relations $\tau =\left( m+3\right) /\left( m+1\right) $ $%
\left[ m\geq 0\right] $ and $\tau =\left( n-4\right) /\left( n-2\right) $ $%
\left[ n\geq 2\right] $, respectively. In Eqs.~(\ref{2}) and (\ref{3}), $%
\Gamma \lbrack .]$ is the Gamma function and $\mathcal{B}_{\tau }$ is given
by
\begin{equation}
\mathcal{B}_{\tau }=\frac{1}{(3-\tau )\,\sigma ^{2} _\tau },  \label{4}
\end{equation}%
where $\sigma _\tau $ is the width of the PDF. For $\tau <5/3$ the width and the
standard deviation, $\sigma $, are naturally related by,
\begin{equation}
(5-3\tau )\,\sigma ^{2}=(3-\tau )\,\sigma ^{2} _\tau.  \label{5}
\end{equation}%

\begin{figure*}[t]
\begin{center}
\vspace{1.0cm}
\includegraphics[width=0.4\textwidth,angle=0]{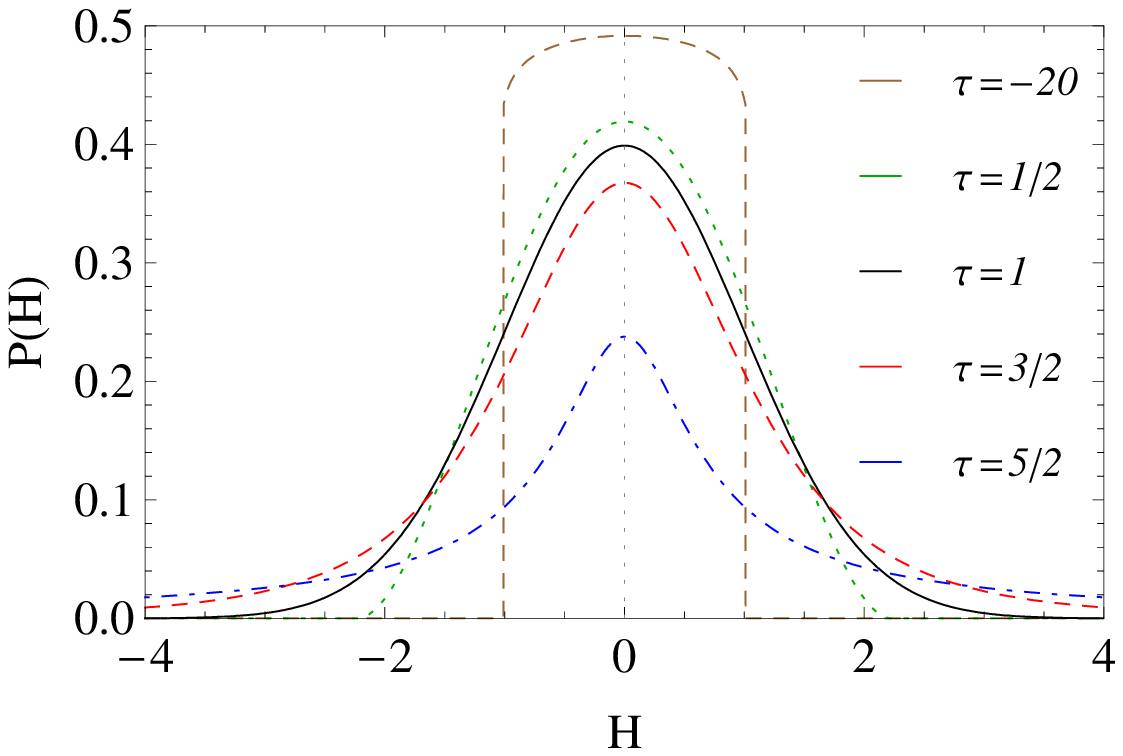}
\hspace{0.5cm}
\includegraphics[width=0.4\textwidth,angle=0]{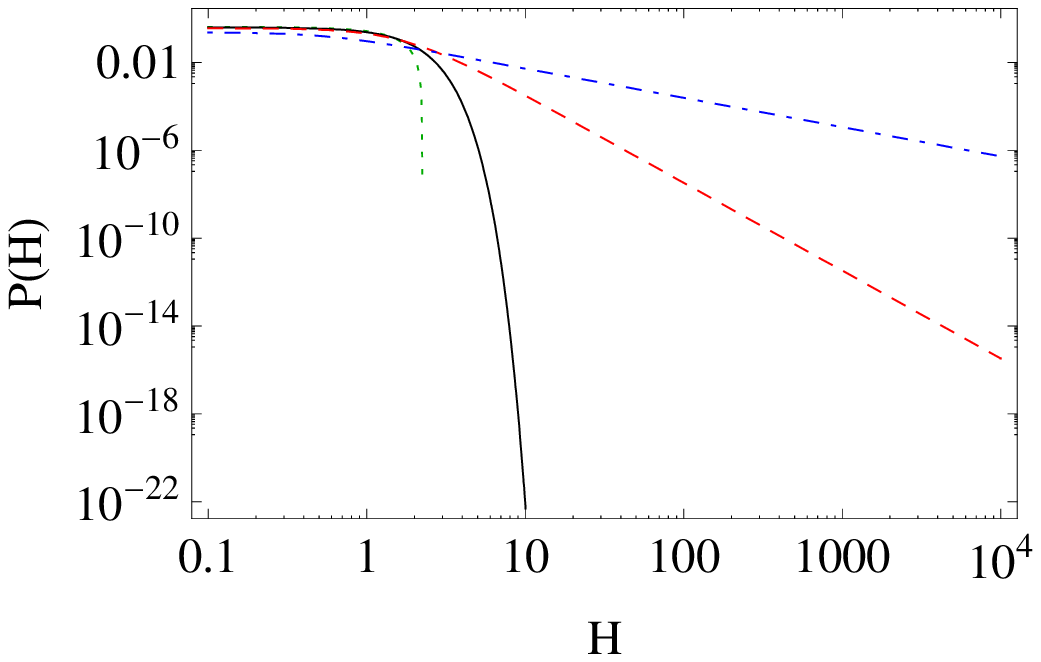}
\end{center}
\caption{(Color online) Random-field probability distributions for some
values of the parameter $\protect\tau $ (from bottom to top: $\protect\tau %
=5/2,3/2,1,1/2$ and $-20$), in the normal (left panel) and log-log scale
(right panel). We have used $\protect\sigma _\tau =1$ in all cases.}
\label{Fig1}
\end{figure*}

In Fig. \ref{Fig1}, we show the PDFs (\ref{2}) and (\ref{3}) for some values of $\tau$, in the linear-linear (left panel) and log-log scales (right panel). We can see the power-law behaviour for $\tau>1$.

For the sake of simplicity, we shall consider $J=1$ hereinafter. We studied systems of $L=6, 8, 10, 12$ and $16$,  with periodic boundary conditions and a random initial configuration of the spins by means of Monte Carlo simulations. The algorithm is as follows: a configuration of the random field $\{H\}$ is generated at the beginning of the simulation according to the corresponding probability $P_{i,s}(H)$, and remains frozen during the whole dynamics; then, every lattice site is visited and a spin flip occurs according to the standard Metropolis rule. The results for all values of the magnetic field parameters $\tau$ and $\sigma_{\tau}$ show that finite-size effects are less-pronounced for $L\geq 10$. We have used $2\times 10^{6}$ MC steps to equilibrate the system and $10^{6}$ MC steps for averaging. Notice that these averages are MC or time averages. Nonetheless, in systems with quenched disorder we need to average over random-field realizations~\cite{binder_young}, \emph{i.e.}, a certain quantity $A$ has the mean value given by,
\begin{equation}\label{6}
\langle A \rangle = \frac{1}{n_{r}}\sum_{i=1}^{n_{r}}\bar{A}_{i} ,
\end{equation}
where $\bar{A}_{i}$ stands for MC average of the \emph{i}-th sample and $n_{r}$ is the number of random-field realizations.
Thus, in addition to the MC steps used for time averaging, we have used number of samples than went up to $2000$ samples of the distributions
$P_{i,s}(H)$, although we afterwards statistically verified by means of the resampling procedure~\cite{jack1} that this number can be substantially
diminished without jeopardizing the results regarding the analysis of the critical nature and frontiers. However, the reader must be aware that for a survey on the universality class of the problem, which is defined by the set of critical exponents, a whopping number of samples is mandatory.

To obtain the quenched random fields we have used standard procedures for generating $t$-Student and $r$-distribution variables~\cite{trgen}.


\section{Numerical Analysis}

As previously discussed~\cite{part1}, the PDFs of Eqs.~(\ref{2})~and~(\ref{3}) may be \textit{platykurtic}, for $\tau<1$, or \textit{leptokurtic}, for $\tau>1$. To study the critical properties of the model we performed MC simulations for $\tau$=$-\infty$, 0.2, 3/2, 5/3 and 2.0, in the range $0.05<T<5.0$ and $0.1\leq\sigma_{\tau}\leq 2.4$. Notice that for $\tau>5/3$, we must use the PDF width $\sigma_{\tau}$ instead of the standard deviation $\sigma$. Next, we will present our results for the order parameter $\langle m\rangle$, the magnetization per spin, and the magnetic susceptibility $\chi$. The latter is obtained from the simulations by making use of the fluctuation-dissipation relation,
\begin{eqnarray}\label{7}
\chi & = & \frac{\langle m^{2}\rangle-\langle m\rangle^{2}}{kT}
\end{eqnarray}

\begin{figure}[t]
\begin{center}
\vspace{1.0cm}
\includegraphics[width=0.4\textwidth,angle=0]{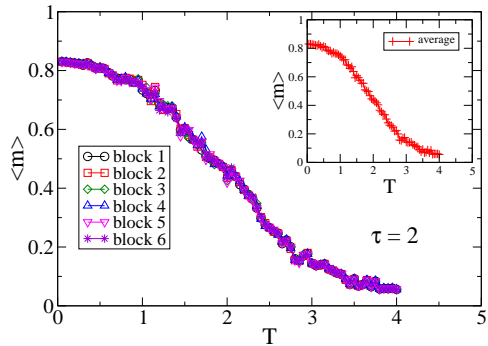}
\end{center}
\caption{(Color online) Magnetization per spin as a function of temperature for $\tau=2$, $\sigma_{\tau}=1.2$, $L=16$ and some blocks with 200 samples each one. In the inset we show the corresponding average magnetization per spin, considering all the 2000 samples.}
\label{Fig_blocks}
\end{figure}

\noindent
where $k$ is the Boltzmann constant (we set $k=1$) and $\langle \; \rangle$ stands for averages given by Eq. (\ref{6}). Before moving forwards we discuss the choice of our sampling. To that aim, we considered a quite broad distribution, $\tau =2$ which is equivalent to the Lorentz distribution, and performed a total of 2000 samples in a system with $L = 16$ on which we have applied a jackknife-like procedure, which belongs to the established class of resampling methods in Statistics~\cite{jackbook}. Explicitly, we regrouped our magnetization data into smaller groups. The smallest group size we considered was 200 elements. Already at this group size we could achieve statistical significance in Student $t$-tests for a $p$-value equal to $0.05$ when we appraised the likelihood of having groups of samples of equal and unequal size (with non-coincident elements obtained from the same $\tau$, $\sigma_{\tau}$ and $T$ parameters) with the same statistical moments. Bear in mind that if we consider 200 samples of a $L=8$ system, we are actually taking into reckoning over than $10^5$ generated quenched random fields, which are sufficient to have a worthy description of the fat tail behaviour, namely the effect of very unlikely random fields. Such a validity can still be intuitively grasped when we plot the magnetization curve considering different group sizes in Fig. \ref{Fig_blocks}. These data are results for simulations for $\tau=2$, $L=16$ and different sub-samples of 200 samples each. It is evident the concurrence between the sub-sets. In the inset of the same figure we plot the mean magnetization averaged over all the 2000 samples which presents the same behaviour as the smaller sets. Once more, we would like to emphasize that we are not setting our sights on the critical universality of the system, but in the definition of critical regions instead, as done in previous cases (see e.g. Ref.~\cite{itakura}).

As made in the mean-field case~\cite{part1}, we will study the platykurtic ($\tau <1$) and the leptokurtic ($\tau >1$) cases separately.

\subsection{Platykurtic case: $\protect\tau <1$}

\begin{figure*}[t]
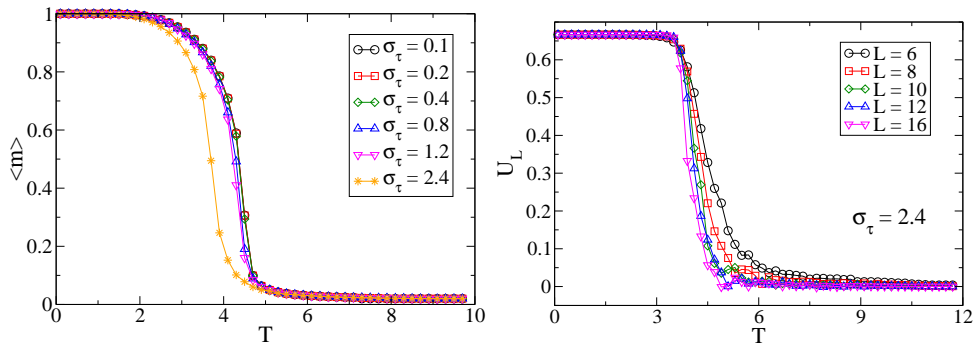

\begin{center}
\vspace{1.0cm}
\includegraphics[width=0.4\textwidth,angle=0]{fig3a.eps}
\includegraphics[width=0.4\textwidth,angle=0]{fig3b.eps}
\end{center}
\caption{(Color online) Magnetization per spin as a function of temperature for a platykurtic case corresponding to the uniform distribution with $L=16$ and typical values of $\sigma_{\tau}$, showing that ferromagnetic order occurs at finite temperatures. For values of $\sigma_{\tau}$ at least up to 2.4, we observe a continuous transition between the ordered and the disordered phases (left side). Observe the small variation of the transition temperatures for increasing values of $\sigma_{\tau}$ as it happens in the mean-field limit~\cite{part1}. It is also shown the Binder cumulant for $\sigma_{\tau}=2.4$ and different lattice sizes (right side). Notice the crossing of the curves at a critical temperature near $T\cong 3.5$.}
\label{Figunif}
\end{figure*}

We start presenting the results for the lower limiting platykurtic case, i.e., the uniform distribution corresponding to $\tau \rightarrow \infty$ (see Fig. \ref{Figunif}). The numerical results suggest that the transition temperature is not much affected by the values of $\sigma_{\tau}$, in agreement with mean-field predictions~\cite{part1}. In addition, plotting the data for the Binder cumulant (left panel of Fig. \ref{Figunif})), obtained by
\begin{equation}
U_{L}=1-\frac{\langle m^{4}\rangle}{3\langle m^{2}\rangle^{2}}~,
\end{equation}
\noindent
we can observe the signature of a phase transition, i.e., the crossing of the curves for different lattice sizes at the critical temperature. For 3D, comparing the latter results with mean-field estimates, we are still able to perceive a phase transition for widths of the distribution greater than 1, which is the maximum value of $\sigma_{\tau}$ wherein symmetry-breaking exists in the uniformly distributed external random-field case.

\begin{figure*}[tbh]
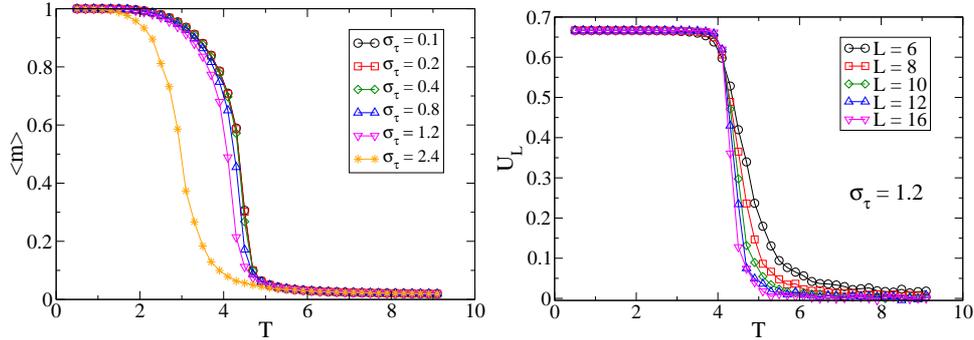

\begin{center}
\vspace{1.0cm}
\includegraphics[width=0.4\textwidth,angle=0]{fig4a.eps}
\includegraphics[width=0.4\textwidth,angle=0]{fig4b.eps}
\end{center}
\caption{(Color online) Magnetization as a function of temperature for another platykurtic case ($\tau=0.2$) and typical values of $\sigma_{\tau}$ (left side). It is also shown the Binder cumulant for $\sigma_{\tau} = 1.2$ and several lattice sizes (left panel), where we can see that the critical temperature is located somewhere near $T\cong 4.1$.}
\label{figt02}
\end{figure*}

In Fig.~\ref{figt02}, we present our results for the magnetization per spin versus the temperature for $\tau=0.2$, $L=16$ and typical values of $\sigma_{\tau}$. One can observe that ferromagnetic order occurs at low temperatures. We have continuous phase transitions between the ordered ferromagnetic phase ($m\to 1$) and the disordered paramagnetic one ($m\to 0$) for all studied values of $\sigma_{\tau}$. In the right panel of the same figure, we depict the scale-behaviour of the Binder cumulant for a particular value of width, $\sigma_{\tau} = 1.2$. Collating both platykurtic cases we observe that the critical temperature wanes as the distribution approaches the Gaussian which is understandable since by increasing $\tau$ we soar the disorder. As mentioned for the uniform case, the cubic lattice presents a larger interval of values of $\sigma_{\tau} $ in which we can verify a phase transition.

In conclusion, if the random fields are generated by platykurtic distributions ($\tau<1$), the system undergoes a phase transition for all studied values of $\sigma_{\tau}$, with a ferromagnetic phase at low temperatures and a paramagnetic phase at high temperatures. For large values of $\sigma_{\tau}$, the system is in the paramagnetic phase for all temperatures. In addition, the critical temperatures are not much affected by increasing values of $\tau$.


\subsection{Leptokurtic case: $\protect\tau >1$}

\begin{figure}[t]
\begin{center}
\vspace{1.0cm}
\includegraphics[width=0.4\textwidth,angle=0]{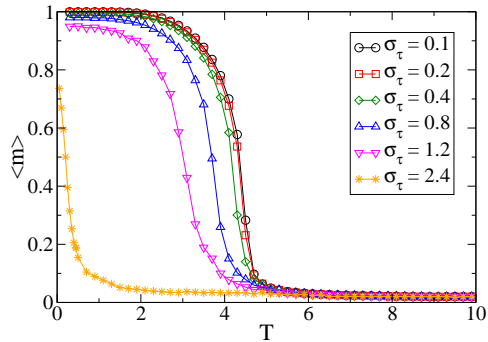}
\end{center}
\caption{(Color online) Magnetization as a function of temperature for the leptokurtic case ($\tau=5/3$), $L=16$ and typical values of $\sigma_{\tau}$. The results suggest that for $\sigma_{\tau}>2.4$ the system is in the paramagnetic phase.}
\label{figt53}
\end{figure}

\begin{figure*}[t]
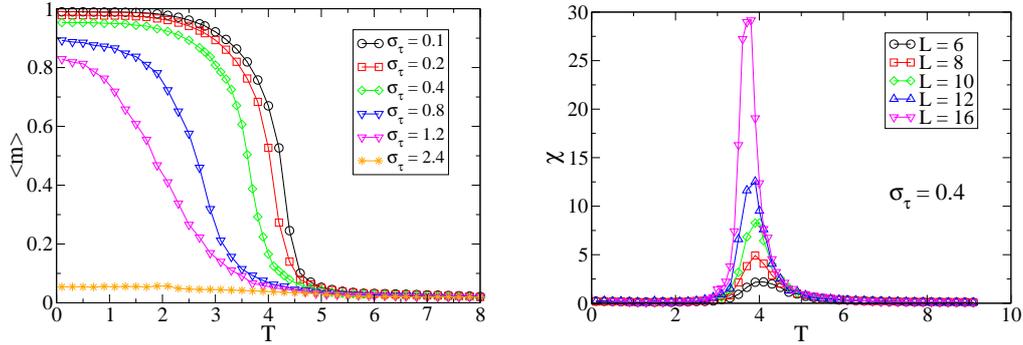

\begin{center}
\vspace{1.0cm}
\includegraphics[width=0.4\textwidth,angle=0]{fig6a.eps}
\hspace{0.5cm}
\includegraphics[width=0.4\textwidth,angle=0]{fig6b.eps}
\end{center}
\caption{(Color online) Magnetization as a function of temperature for another leptokurtic case case ($\tau=2$) $\sigma_{\tau} = 1.2$ and several values of the system size (left panel); Magnetic susceptibility as a function of temperature for another leptokurtic case case ($\tau=2$) $\sigma_{\tau} = 0.4$ and several values of the system size (right panel).}
\label{figt2}
\end{figure*}

The picture remains very much the same when we study systems presenting fat tailed behaviour in the probability density function of the quenched random fields. Namely, we have found evidence for a continuous phase transition for all the values of the parameter $\tau $ we simulated. However, the critical temperatures are more sensitive to variations in $\sigma_{\tau}$ than the platykurtic region, in agreement with mean-field predictions~\cite{part1}. Concomitantly, we have understood a decrease of the maximum value of $\sigma_{\tau}$, $\sigma_{\tau}^{\max}$, for which it is possible to identify the existence of a ferro-paramagnetic transition at finite temperature. Once again, it is straightforward to verify that the greater $\tau $, the smaller $\sigma_{\tau}^{\max}$, since an augment of the value of $\tau $ encloses an increase of the disorder. For instance, comparing the results for $\tau = 5/3$ and $\tau = 2$, we notice that fixing $\sigma_{\tau} = 2.4$, we distinguish a finite temperature phase transition for the former $\tau $ while we cannot descry the existence of critical behaviour at finite temperature for the latter (see Figs. \ref{figt53} and \ref{figt2}). Comparing with high-dimensional result (mean field), we have the limit width at $T=0$ for $\sigma_{\tau = 5/3} = 1/\sqrt{2}$ and $\sigma_{\tau = 2} = 2/ \pi $~\cite{part1}. We must note that the $\tau = 2$ corresponds to the sub-domain of values of the tail index we obtain a divergent free energy at $T= 0$ in the mean-field limit was calculated.

\begin{figure}[t]
\begin{center}
\vspace{1.0cm}
\includegraphics[width=0.4\textwidth,angle=0]{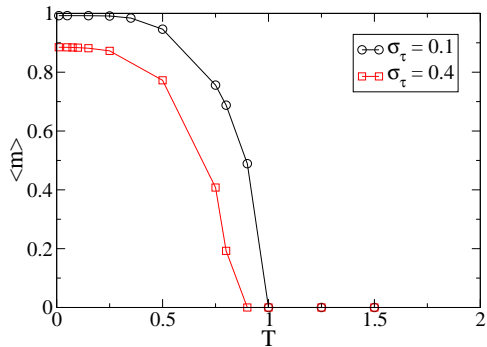}
\end{center}
\caption{(Color online) Mean-field magnetization as a function of temperature for $\tau=2$ and $\sigma_{2} = 0.1$ and $0.4$. We can observe in this high-dimensional limit the same behaviour at low temperatures presented by the three-dimensional case. The points were obtained numerically and the line is presented as guide to the eye.}
\label{figt2_mf}
\end{figure}

One of the points that called our attention is the fact that for large values of the disorder, particularly when there is an obvious tail effect, the magnetization we found at low temperatures is significantly different from 1. In order to check this result we made use of the mean-field treatment. Explicitly, we compared the numerical results from our Monte Carlo simulations with the solutions to the mean-field equation~\cite{part1},
\begin{equation}
m = \int_{-\infty}^{+\infty}{\rm d}H\;P_s(H)\tanh \beta (J \, m + H) ~.
\end{equation}
The results of two cases are depicted in Fig. \ref{figt2_mf}. For a very small value of $\sigma_{\tau}$ (0.1), the low-temperature magnetization per spin is equal to 1, as in the three-dimensional case, but it is visible that for $\sigma_{\tau}=0.4$ we can have $m<0.9$ at low temperatures. The integration was carried out using the global adaptive strategy algorithm~\cite{kromer}, which for the two cases exhibited has an maximum error of $0.4 \% $ in the numerical integration. Moreover, for large values of the temperature, which imply a paramagnetic phase, we obtained values of $| m | \cong 0.003$, which swayed us to consider that this effect as a signature of the system, particularly of the tail.

\begin{figure}[t]
\begin{center}
\vspace{1.0cm}
\includegraphics[width=0.4\textwidth,angle=0]{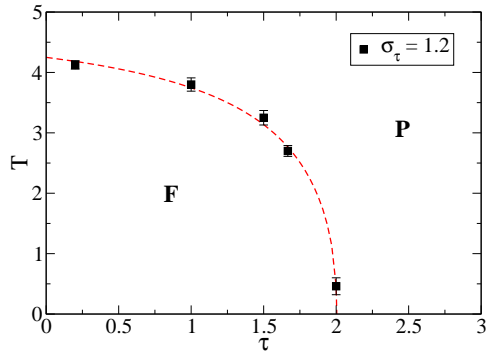}
\end{center}
\caption{(Color online) Sketch of the phase diagram of the model in the plane Temperature versus $\tau$, for $\sigma_{\tau}=1.2$, separating the Ferromagnetic (\textbf{F}) and Paramagnetic (\textbf{P}) phases. The points are Monte Carlo estimates of the critical temperatures and and the line is the fit using Eq. (\ref{fit}).}
\label{fig_ph_diag}
\end{figure}

With these and other results we obtained, it is possible to sketched the diagram of the temperature $T$ versus $\tau $ for a fixed value $\sigma_{\tau}=1.2$ (see Fig. \ref{fig_ph_diag}). The squares are MC estimates of the critical temperatures, and we have considered the critical temperature of references \cite{machta00,itakura} for the Gaussian case ($\tau=1$). The computational results are in good agreement with the heuristic expression,
\begin{equation}\label{fit}
T(\tau)=a\left[1-(\tau-\tau_{c})^{-1}\right]^{-b} ~,
\end{equation}
\noindent
when $a=5.08 \pm 0.16$, $b=0.44 \pm 0.04$ and $\tau_{c}=2.004 \pm 0.002$ ($R^2 = 0.998$ and $\chi ^2 /n = 0.007$). This curve is shown in Fig. \ref{fig_ph_diag} as a dashed line. Observe that for $\tau>1$ the critical temperatures decrease rapidly with increasing values of $\tau$, but they do not decrease significantly for $\tau<1$, as above-discussed. The validity of the adjustment can be tested by comparing the asymptotic limit of our ansatz~(\ref{fit}), which is equal to $a$, with the critical temperature of the uniform distribution case, $T_c = 4.41$. This approximately corresponds to a $ 10 \% $ difference (considering error) which is acceptable taking into account the reduced number of parameters we assumed so that the Akaike information criterion is implicitly minimized~\cite{akaike}.

\section{Remarks and Outlook}

In this work, we have presented a preliminary survey on both the existence and the nature of critical behaviour of the random-field Ising model on cubic lattices with nearest-neighbour interactions by means of Monte Carlo simulations based on the standard Metropolis algorithm. In order to generate the quenched random fields, we have considered a family of continuous probability density functions, defined by a parameter $\tau $ comprising the $r$-distribution, for $\tau <1$, and the Student-$t$, for $\tau >1$. This set of distributions allowed us to study a larger variety of cases that go from the uniform distribution to fat-tailed distributions with the same asymptotic regime as the L\'{e}vy distribution that were already used in the study of vitreous systems~\cite{cizeaubouchaud}, and for which it was recently found that the local fields follow non-Gaussian distributions~\cite{levyglass,mezardepl}. These models are considered relevant to the description of spin-glasses with Ruderman-Kittel-Kasuya-Yosida interaction and spin glasses exhibiting a wide hierarchy of coupling intensities as well. The current results show that in the cubic lattice, this RFIM model presents many of the traits found for the mean-field case~\cite{part1}. Namely, based on the analysis of different quantities such as the order parameter, response function and Binder cumulant, we have found evidence for the existence of continuous phase transitions for all values of $\tau $ with the critical temperature, $T_c$, depending on the characteristic width of the distribution, $\sigma _\tau $, and the tail index, $\tau $, itself. As expected, the spell of $\sigma _\tau $ in which we can detect a ferro-paramagnetic phase transition is greater than the region found in the mean-field approach. Although we might be tempted to think that the parameter $\tau $ represents just a supplementary way of tuning the non-thermal disorder introduced in the system, we must take into account results obtained at $T=0$ where the thermal disorder is quelled. Our previous calculations have shown quite different behaviour of thermodynamical quantities depending on the parameter $\tau $, particularly for distributions fatter than the Lorentzian. We expect that the analysis of the ground state together with upgraded simulations we are also carrying out manage to shed further light on the actual impact of the tails.

\ack{
We would like to thank M.~Azeinman for exchange of correspondence and I.~Giardina for clarifying comments. We would also like to acknowledge the collaboration of D.O.~Soares-Pinto during the embryonic phase of this work.
NC thanks the financial support from the Brazilian funding agency CNPq.}

\end{document}